\documentclass[cameraready]{Interspeech}
\title{NV-Bench: Benchmark of Nonverbal Vocalization Synthesis for Expressive Text-to-Speech Generation}
\author{Qinke}{Ni}
\author{Huan}{Liao}
\author{Dekun}{Chen}
\author{Yuxiang}{Wang}
\author{Zhizheng}{Wu}
\address{
    The Chinese University of Hong Kong, Shenzhen
}

\email{qinkeni@link.cuhk.edu.cn, huanliao@link.cuhk.edu.cn, dekunchen@link.cuhk.edu.cn, yuxiangwang1@link.cuhk.edu.cn, wuzhizheng@cuhk.edu.cn}


\usepackage{comment}
\usepackage{booktabs}
\usepackage{pifont}
\usepackage{xcolor}
\usepackage{tabularx}
\usepackage{array}
\usepackage{makecell} % for line breaks in header
\usepackage{multirow}
\usepackage{colortbl} % 

\newcommand{\cmark}{\ding{51}} % ✓
\newcommand{\xmark}{\ding{55}} % ✗
\definecolor{grayrow}{gray}{0.92}

\begin{document}

\maketitle

\begin{abstract}
  While recent text-to-speech (TTS) systems increasingly integrate nonverbal vocalizations (NVs), their evaluations lack standardized metrics and reliable ground-truth references. To bridge this gap, we propose NV-Bench, the first benchmark grounded in a functional taxonomy that treats NVs as communicative acts rather than acoustic artifacts. NV-Bench comprises 1,651 multi-lingual, in-the-wild utterances with paired human reference audio, balanced across 14 NV categories. We introduce a dual-dimensional evaluation protocol: (1) Instruction Alignment, utilizing the proposed paralinguistic character error rate (PCER) to assess controllability, (2) Acoustic Fidelity, measuring the distributional gap to real recordings to assess acoustic realism. We evaluate diverse TTS models and develop two baselines. Experimental results demonstrate a strong correlation between our objective metrics and human perception, establishing NV-Bench as a standardized evaluation framework.
\end{abstract}

\begin{figure*}[t]
  \centering
  \includegraphics[width=\textwidth]{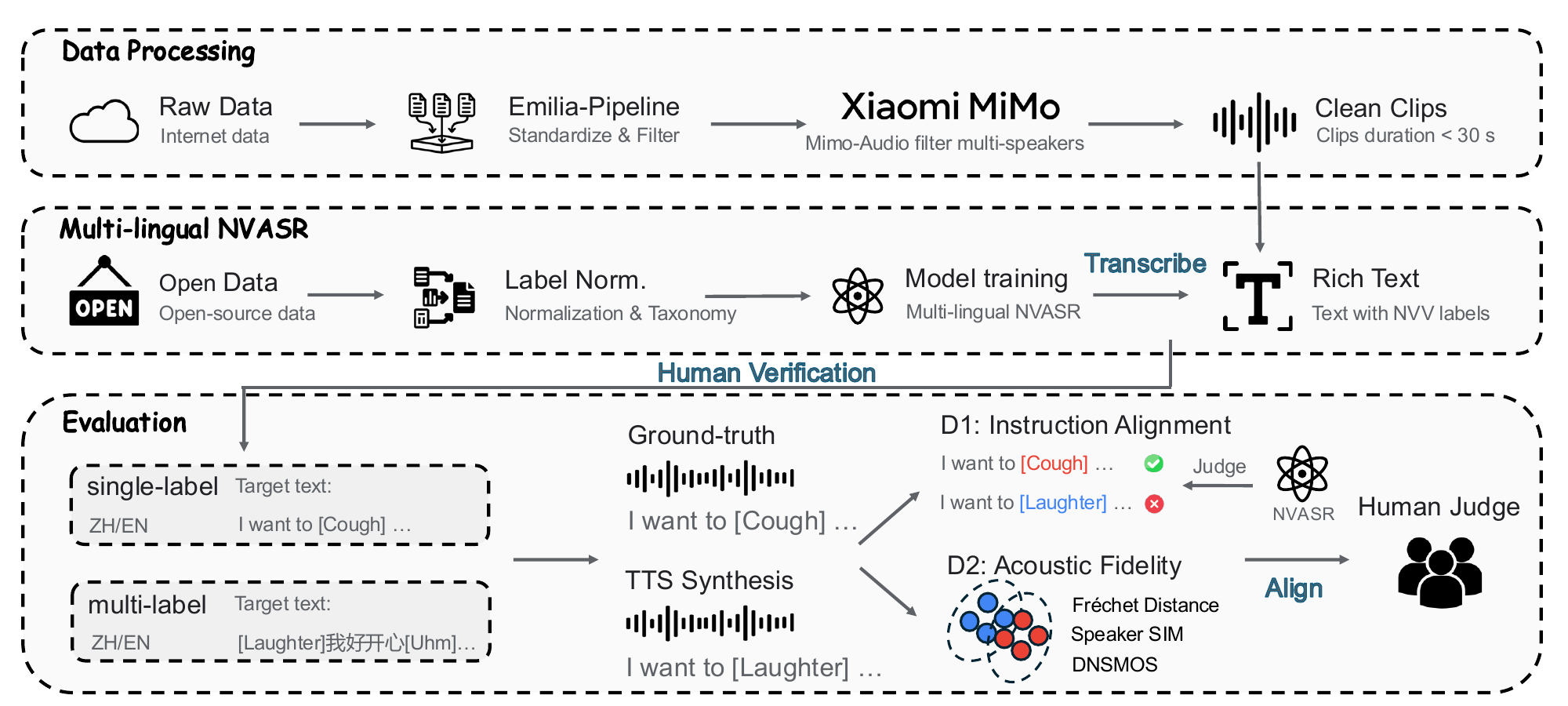}
  \caption{\textbf{Overview of the NV-Bench.} (1) Data Processing: Raw audio is filtered using the Emilia-Pipeline and MiMo-Audio. (2) Multi-lingual NVASR: We train a multi-lingual NVASR model on open-source data with a unified label taxonomy. (3) Evaluation: After human verification, the benchmark is evaluated in instruction alignment and acoustic fidelity dimensions and human subjective ratings.}
  \label{fig:overview}
\end{figure*}

\section{Introduction}

Recent expressive text-to-speech (TTS) models \cite{wang2024maskgct,chen2025f5,yan2025step} increasingly integrate nonverbal vocalizations (NVs) to enhance human-like communication. Current methods incorporate NVs either as discrete tokens (e.g., NVTTS \cite{borisov2025nonverbaltts}) or overlapping layers (e.g., CapSpeech \cite{wang2025capspeech}). However, these approaches predominantly treat NVs as generic "sound effects" annexed to linguistic content. This mechanistic perspective overlooks their fundamental nature: NVs are not merely acoustic textures, but intrinsically communicative acts conveying physiological states, emotions, and interactional intent. Advancing this field requires moving beyond checking for acoustic presence to evaluating pragmatic appropriateness in context.

To systematically model these phenomena, we adopt the functional taxonomy of Batliner et al.~\cite{batliner2025non}, categorizing NVs into three levels crucial for expressive TTS: (1) \textbf{Vegetative sounds} encompass biological reflexes such as breathing and coughing, which ground generated speech in physical realism. (2) \textbf{Affect bursts} consist of valenced vocalizations that succinctly convey the speaker's emotional state or instantaneous reactions. (3) \textbf{Conversational grunts} function as interaction-management cues, including filled pauses and prosodic particles that disambiguate communicative intent (e.g., confirmation or hesitation).
This taxonomy reveals NVs are not merely acoustic events, they encompass a spectrum of non-lexical, pragmatically charged interjections essential for conveying emotion and discourse management.

Motivated by the need for expressive systems, several large-scale corpora with NVs have been proposed (Table \ref{tab:nonverbal-datasets}), including Emilia-NV \cite{liao2025nvspeech}, SMIIP-NV \cite{wu2025smiip}, NVTTS \cite{borisov2025nonverbaltts}, DisfluencySpeech \cite{wang2024disfluencyspeechsinglespeakerconversational}, NonverbalSpeech (NVS) \cite{ye2025scalable}, SynParaSpeech \cite{bai2025synparaspeech}. But scaling training data does not yield a reliable evaluation standard. Beyond ambiguous task definitions, current evaluation practices rely on internal testsets or text-rewritten references rather than authentically paired human speech data. Without ground-truth (GT) NV recordings, evaluation tends to collapse to coarse checks (e.g., event presence/absence), making it impossible to quantify the gap to real recordings. Furthermore, given that NV events are long-tailed, imbalanced testsets can bias aggregated metrics and hinder fair diagnosis.
 
To address these gaps, we introduce \textbf{NV-Bench}\footnote{Demo page: \url{https://nvbench.github.io}}, a comprehensive benchmark for NV-capable TTS. NV-Bench provides a public multi-lingual testset comprising 1,651 utterances which are curated from online audiovisual media published in 2025 to minimize the potential data leakage. To ensure fair comparison, we partition the benchmark into a strictly balanced single-label subset (50 utterances per category) and a relatively balanced multi-label subset with 14 NV types. All test samples contain in-the-wild NVs paired with GT audio. This paired design enables the reproducible assessment of instruction alignment via character error rate (CER) and its variants and acoustic fidelity via speaker similarity and fréchet distance (FD).

Our main contributions are threefold: 
(1) We propose \textbf{NV-Bench}, the first comprehensive evaluation framework for NV-capable TTS, featuring a public, in-the-wild dataset with paired human GT audio. 
(2) We establish a standardized and distributionally balanced evaluation protocol that supports fair and reproducible model comparisons. 
(3) We conduct extensive benchmarking of state-of-the-art (SOTA) TTS models, revealing their controllability, intelligibility and acoustic fidelity.

\begin{table}[t]
  \centering
  \small
  \caption{Comparison with nonverbal vocalizations datasets.}
  \label{tab:nonverbal-datasets}
  \begin{tabularx}{\columnwidth}{p{2.2cm} c c c c}
    \toprule
    \textbf{Dataset} & \textbf{Lang.} & \textbf{Testset} & \textbf{Balance} & \textbf{Prompt} \\
    \midrule
    SynParaSpeech~\cite{bai2025synparaspeech} & zh & \xmark & -- & -- \\
    NVS~\cite{ye2025scalable} & zh/en & \xmark & -- & --  \\
    Emilia-NV~\cite{liao2025nvspeech} & zh & \xmark & -- & -- \\
    NVTTS~\cite{borisov2025nonverbaltts} & en & \cmark & \xmark & \xmark  \\
    SMIIP-NV~\cite{wu2025smiip} & zh  & \cmark & \xmark & \xmark \\
    \midrule
    \textbf{NV-Bench} & \textbf{zh/en} & \cmark  & \cmark & \cmark \\
    \bottomrule
  \end{tabularx}
\end{table}

\section{Methods}

To construct \textbf{NV-Bench}, we introduce a two-phase pipeline balancing acoustic diversity and label accuracy: (1) developing a robust multi-lingual NVASR to provide high-quality, event-aware transcriptions, (2) curating in-the-wild audio through a rigorous filter that culminates in human-verified GT.

\subsection{Multi-lingual NVASR}
\label{sec:nvasr}

To facilitate efficient benchmark construction, we develop a robust multi-lingual NV-capable automatic speech recognition (NVASR) model. Following the methodology introduced in~\cite{liao2025nvspeech}, we finetune the SenseVoice-Small~\cite{an2024funaudiollm} architecture and extend the framework to support multilingual settings.

\subsubsection{Model architecture}
We select SenseVoice-Small~\cite{an2024funaudiollm} as our base model due to its pre-training on diverse audio understanding tasks, which allows it to capture rich features. The model is optimized by minimizing the connectionist temporal classification (CTC) loss~\cite{graves2006connectionist}:
\begin{equation}
\mathcal{L}_{\text{CTC}} = -\ln \sum_{\pi \in \mathcal{B}^{-1}(\mathbf{y})} P(\pi \mid \mathbf{x})
\end{equation}
where $\mathbf{x}$ represents the input acoustic features, $\mathbf{y}$ is the target sequence, $\mathcal{B}^{-1}(\mathbf{y})$ denotes all valid CTC alignment paths for $\mathbf{y}$.

\subsubsection{Data construction and label normalization}
\label{label_norm}
To ensure broad generalization, we consolidate a comprehensive training corpus: Emilia-NV~\cite{liao2025nvspeech}, NVTTS~\cite{borisov2025nonverbaltts}, DisfluencySpeech~\cite{wang2024disfluencyspeechsinglespeakerconversational}, NVS~\cite{ye2025scalable}, SMIIP-NV~\cite{wu2025smiip}, and MNV-17~\cite{mai2025mnv}.

To address the heterogeneity of source labels, we implement a 
systematic normalization process. We map non-speech labels to Level 3 of the AudioSet Ontology~\cite{gemmeke2017audio}, preserving distinct events like ``[Laughter]''. We adopt the Emilia-NV taxonomy but conduct targeted manual annotations on the English subsets of NVTTS and DisfluencySpeech. This step is critical to distinguish nuanced pragmatic functions, such as ``[Question-huh]'' that were previously absent in English datasets. The final unified taxonomy is detailed in Table~\ref{tab:nvasr-label-inventory}.

\begin{table*}[t]
  \centering
  \caption{Unified label inventory categorized by communicative function.}
  \label{tab:nvasr-label-inventory}
  \small
  \begin{tabularx}{\textwidth}{l >{\raggedright\arraybackslash}p{3.5cm} >{\raggedright\arraybackslash}X >{\raggedright\arraybackslash}X}
    \toprule
    \textbf{Language} 
    & \textbf{Vegetative Sounds} 
    & \textbf{Affect Bursts} 
    & \textbf{Conversational Grunts} \\
    \midrule
    Mandarin 
    & Breathing, Cough, Sigh
    & Laughter, Surprise-ah, Surprise-oh, Dissatisfaction-hnn 
    & Uhm, Confirmation-en, Question-ei, Question-ah, Question-en, Question-oh \\
    \addlinespace
    English 
    & Breathing, Cough, Sigh
    & Laughter, Surprise-oh
    & Uhm, Question-huh \\
    \bottomrule
  \end{tabularx}
\end{table*}

\subsection{Benchmark dataset construction}

Constructing a reliable NV benchmark requires balancing in-the-wild realism with label distribution. We curate our dataset entirely from diverse web-sourced audio using a sophisticated filtering pipeline, as shown in Figure \ref{fig:overview}.

\subsubsection{Data collection} 
NV events naturally exhibit a long-tail distribution. To ensure adequate coverage across our taxonomy, we crawled a massive collection of audiovisual media uploaded within the most recent year. From a raw pool of 565,316 audio clips ($\approx$ 1,560 hours), we filtered for candidate segments containing target NVs.

\subsubsection{Filtering pipeline}
To guarantee acoustic fidelity and speaker purity, we subject the candidate segments to a rigorous refinement process:

\textbf{Audio Standardization.} We first utilize the Emilia-Pipeline \cite{he2024emilia} for initial audio standardization and source separation. While this includes speaker diarization, multi-speaker segments often persist.

\textbf{Single-Speaker Verification.} To eliminate residual multi-speaker artifacts, we deploy MiMo-Audio-7B-Instruct \cite{coreteam2025mimoaudio}, a SOTA audio large language model (LLM), prompting it to detect subtle speaker overlaps that escaped initial diarization. Only confirmed clean, single-speaker utterances are retained.

\textbf{Human Verification.} In the final stage, ten expert annotators review and correct the NVASR-generated transcripts, validating the pragmatic appropriateness of NV labels. To ensure annotation consistency, 5\% of the data was cross-annotated, achieving a high Cohen's kappa above 0.85. This process yields a dataset of 1,651 prompt and GT pairs (7.9 hours), which are standardized to MP3 format at a 24\,kHz sampling rate.

\section{NV-Bench}

Despite the proliferation of NV-capable TTS systems, the field lacks a standardized benchmark to disentangle two distinct failure modes: (i) failure to generate the intended event, and (ii) generation of low-quality or unnatural audio. To address this, NV-Bench evaluates systems across two dimensions:

\textbf{Instruction Alignment:} Evaluates the model's ability to strictly follow textual prompts. Specifically, we assess whether the system generates target NV events at precise linguistic positions without omissions or hallucinations, serving as a proxy for robust text-to-speech alignment.

\textbf{Acoustic Fidelity:} Assesses realism relative to real recordings. We quantify the distributional gap, timbre consistency and their perception quality to evaluate their acoustic fidelity.

To probe these capabilities under varying complexity, NV-Bench is structured into two multi-lingual subsets:

\textbf{Single-label Subset:} A strictly balanced testset where each utterance contains exactly one NV event (50 samples per category, 650 Mandarin and 350 English utterances in total), which isolates fundamental generation capabilities.

\textbf{Multi-label Subset:} A more challenging set containing utterances with multiple (2+) NV events to test robustness under dense paralinguistic conditions. Acknowledging the long-tailed nature of natural co-occurrences, we ensure relative balance (Mandarin: 41--91, English: 75--112 samples per label), providing sufficient coverage for all event types while reflecting real-world complexity.

\begin{table}[tb]
\centering
\small
\caption{Comparison of CER and OCER (\%) across testsets, where bracketed values are OCER and other values are CER.}
\label{tab:combined_asr_results}
\begin{tabular}{l|cc|c}
\toprule
Dataset & SV & Qwen2.5-Omni & \textbf{NVASR} \\
\midrule
WS-net & 5.77 & 20.14 & \textbf{5.55} \\
LS-other & 12.79 & 23.35 & \textbf{9.90} \\
\midrule
SMIIP-NV & 3.12 & 3.59 (4.17) & \textbf{1.29} (\textbf{1.36}) \\
NVTTS & 14.45 & 21.69 (26.95) & \textbf{13.52} (\textbf{16.10}) \\
\bottomrule
\end{tabular}
\end{table}

 % --- TABLE 1: Language Specific Alignment & SIM ---
\begin{table*}[t]
\centering
\caption{Detailed performance on each subsets of NV-Bench. \textbf{Bold} indicates best in the column, \underline{underline} second-best.}
\label{tab:detailed_results}
% \small
\setlength{\tabcolsep}{3.5pt}
\begin{tabular}{l|ccc|cc|ccc|cc}
\toprule
\multirow{3}{*}{\textbf{System}} & \multicolumn{5}{c|}{\textbf{Single-label Subset}} & \multicolumn{5}{c}{\textbf{Multi-label Subset}} \\
\cmidrule(lr){2-6} \cmidrule(lr){7-11}
& \multicolumn{3}{c|}{Alignment} & \multicolumn{2}{c|}{Fidelity} & \multicolumn{3}{c|}{Alignment} & \multicolumn{2}{c}{Fidelity} \\
& CER (\%) & PCER (\%) & OCER (\%) & SIM & DNSMOS & CER (\%) & PCER (\%) & OCER (\%) & SIM & DNSMOS \\
\midrule
\multicolumn{11}{l}{\textit{\textbf{Mandarin}}} \\
\midrule
\rowcolor{gray!20}
GT         & 3.86 & 9.38 & 4.07 & 0.781 & 3.12 & 3.79 & 23.71 & 5.07 & 0.794 & 3.12 \\
Orpheus-TTS         & 11.36 & 88.77 & 13.91 & - & \textbf{3.43} & 19.83 & 84.85 & 24.38 & - & \textbf{3.40} \\
SMIIP-NV-CV2        & 8.80 & 75.64 & 11.34 & 0.719 & 3.22 & 10.66 & 77.20 & 14.79 & 0.715 & 3.07 \\
Emilia-NV-CV2       & 5.05 & 40.00 & 6.64 & 0.740 & 3.21 & 5.54 & 48.74 & 8.09 & 0.746 & 3.24 \\
CosyVoice3  & \underline{3.85} & 57.69 & \underline{5.86} & \underline{0.764} & \underline{3.30} & 
\underline{4.75} & 61.94 & \underline{8.26} & 0.715 & \underline{3.31} \\
\textbf{NV-FlexiVoice}        & 6.98 & \underline{31.08} & 8.15 & 0.748 & 3.22 & 8.20 & \underline{39.37} & 10.39 & \underline{0.750} & 3.07 \\
\textbf{NV-CV3}  & \textbf{3.80} & \textbf{27.69} & \textbf{4.90} & \textbf{0.768} & 3.29 & \textbf{3.44} & \textbf{30.04} & \textbf{4.84}  & \textbf{0.776} & 3.29 \\
\midrule
\multicolumn{11}{l}{\textit{\textbf{English}}} \\
\midrule
\rowcolor{gray!20}
GT         & 6.73 & 8.31 & 6.90 & 0.772 & 3.11 & 7.26 & 21.41 & 8.62 & 0.775 & 3.14 \\
Orpheus-TTS         & 9.03 & 71.92 & 10.63 & - & \textbf{3.33} & 8.68 & 71.46 & 11.89 & - & \textbf{3.34} \\
SMIIP-NV-CV2        & 17.92 & 56.80 & 19.47 & 0.583 & 2.97 & 20.93 & 54.49 & 23.87 & 0.580 & 2.97 \\
Emilia-NV-CV2      & 12.50 & 55.30 & 13.21 & 0.639 & 3.21 & 11.71 & 60.28 & 14.63 & 0.655 & 3.26 \\
CosyVoice3  & \textbf{7.87} & 62.75 & \textbf{9.06} & \textbf{0.701} & \underline{3.27} & \textbf{6.39} & 57.84 & \underline{10.69} & \underline{0.715} & \underline{3.31} \\
\textbf{NV-FlexiVoice}        & 11.88 & \underline{50.43} & 13.21 & 0.685 & 3.15 & 9.60 & \underline{51.32} & 13.76 & 0.708 & 3.07 \\
\textbf{NV-CV3} & \underline{8.33} & \textbf{46.13} & \underline{9.44} & \underline{0.698} & 3.24 & \underline{6.70} & \textbf{47.13} & \textbf{10.10} & \textbf{0.721} & 3.30 \\
\bottomrule
\end{tabular}
\end{table*}

\section{Experiments}

\subsection{Experiments for multi-lingual NVASR}

As the backbone evaluator for NV-Bench, the performance of multi-lingual NVASR needs to be qualified in both standard automatic speech recognition (ASR) testsets and NV testsets.

\subsubsection{Experimental setup and baselines}

We compare our NVASR against two strong baselines:

\textbf{SenseVoice-Small (SV) \cite{an2024funaudiollm}:} The original ASR model before finetuning, as a baseline for general ASR capability.

\textbf{Qwen2.5-Omni \cite{xu2025qwen25omnitechnicalreport}:} A 7B multimodal LLM capable of speech understanding. We use the finetuned checkpoint for NV recognition released in \cite{mai2025mnv}.

We evaluate on four testsets: WenetSpeech test-net (WS-net) \cite{zhang2022wenetspeech} and LibriSpeech test-other (LS-other) \cite{panayotov2015librispeech} for general ASR performance; and SMIIP-NV \cite{wu2025smiip}, NVTTS \cite{borisov2025nonverbaltts} testset for NV-specific performance.

\subsubsection{Evaluation metrics}
\label{asr_eval_metric}

To jointly evaluate linguistic and paralinguistic accuracy, we extend standard CER into Overall CER (OCER):
\begin{equation}
    \text{OCER} = \frac{S + D + I}{N_{\text{text}} + N_{\text{nvv}}} \times 100\%
\end{equation}
where $S, D, I$ represent substitutions, deletions, and insertions computed over the full sequence including NV labels and $N_{\text{text}}, N_{\text{nvv}}$ denote the count of text characters and NV symbols.

\subsubsection{Results}
As detailed in Table~\ref{tab:combined_asr_results}, our multi-lingual NVASR model demonstrates dual-capability. First, it strictly maintains and even marginally improves upon, the high-quality speech transcription of the original SenseVoice model on standard datasets. Second, it significantly outperforms the MNV-17 finetuned Qwen2.5-Omni in accurately detecting and classifying NV labels. Notably, on the SMIIP-NV testset, our NVASR achieves 1.29\% CER and 1.36\% OCER, confirming its reliability as an automated evaluator for NV-Bench.

\begin{table}[t]
\centering
\caption{Evaluation on the full NV-Bench.}
\label{tab:global_fad}
\small
\begin{tabular}{lcccc}
\toprule
\textbf{System} & \textbf{FAD} & \textbf{FD} & \textbf{IMOS} & \textbf{NMOS} \\
\midrule
\rowcolor{gray!20}
GT & - & -  & 4.39 ± 0.18 & 4.39 ± 0.15  \\
Orpheus-TTS & 5.71 & 24.49 & 3.27 ± 0.23 & 3.53 ± 0.22 \\
SMIIP-NV-CV2 & 1.32 & 6.71 & 3.28 ± 0.22 & 3.28 ± 0.19 \\
Emilia-NV-CV2 & 1.08 & 5.57 & 3.89 ± 0.18 & 3.99 ± 0.14 \\
CosyVoice3 & 0.90 & 9.46 & 3.56 ± 0.22 & 3.94 ± 0.20 \\
\textbf{NV-FlexiVoice} & \textbf{0.29} & \textbf{2.72} & \underline{3.94 ± 0.23} & \underline{4.00 ± 0.18} \\
\textbf{NV-CV3} & \underline{0.86} & \underline{3.94} & \textbf{3.95 ± 0.18} & \textbf{4.08 ± 0.16} \\
\bottomrule
\end{tabular}
\end{table}

\subsection{Benchmarking NV-Capable TTS models}

We conduct a comprehensive evaluation on \textbf{NV-Bench} to assess the current performance of NV-capable synthesis in two dimensions: (1) \textbf{Instruction Alignment}, measures prompt controllability and paralinguistic intelligibility; (2) \textbf{Acoustic Fidelity}, evaluates the distributional gap, speaker similarity, perceptual quality to assess the realism of the synthesized audio.

\subsubsection{TTS systems}

\hspace*{\parindent}\textbf{Orpheus-TTS}\footnote{\url{https://orpheustts.net}}: A single-speaker Llama-based 3B model with explicit NV control.

\textbf{SMIIP-NV-CV2 \cite{wu2025smiip} \& Emilia-NV-CV2 \cite{liao2025nvspeech}}: Two variants of the 0.5B CosyVoice2 \cite{du2024cosyvoice}, a zero-shot TTS model, finetuned respectively on the SMIIP-NV and Emilia-NV datasets.

\textbf{CosyVoice3 (CV3) \cite{du2025cosyvoice}}: The foundational zero-shot model, serving as a strong baseline for general speech synthesis.

\textbf{NV-FlexiVoice}: We finetune a 0.5B FlexiVoice model \cite{chen2026flexivoice} pretrained on the Emilia dataset \cite{he2024emilia} without NV events.

\textbf{NV-CV3}: To establish a high-performance reference baseline, we finetune CV3 on a consolidated corpus.

\subsubsection{Experimental setup}

\hspace*{\parindent}\textbf{Training Configuration.} We finetune CV3 and FlexiVoice using a consolidated corpus comprising Emilia-NV, SMIIP-NV, NVTTS, Disfluency, and NVS. Following Section \ref{label_norm}, normalized NV labels were injected into the text as special control symbols. The model was optimized using AdamW \cite{loshchilov2017decoupled} ($lr=1\times 10^{-5}$) on 4 NVIDIA A800 GPUs.

\textbf{Inference Procedures.} All models follow the label normalization protocol defined in Section \ref{label_norm}. For baselines with limited NV support, we evaluate only the intersection of supported events. Unsupported interjections (e.g., [Question-huh]) are mapped to their nearest lexical equivalents with punctuation (e.g., ``huh?'') to approximate the intended pragmatic function.

\subsubsection{Evaluation metrics} 

\hspace*{\parindent}\textbf{Instruction alignment.} We utilize our NVASR (Section \ref{sec:nvasr}) to compute CER, OCER and PCER. To isolate NV generation accuracy, PCER is calculated on extracted NV symbols: $\text{PCER} = (S_{\text{nvv}} + D_{\text{nvv}} + I_{\text{nvv}}) / N_{\text{nvv}}$, where $S_{\text{nvv}}, D_{\text{nvv}}, I_{\text{nvv}}$ are edit operations on NVs and $N_{\text{nvv}}$ is the target NV count.

\textbf{Acoustic Fidelity.} Perceptual quality and timbre consistency are measured using DNSMOS and WavLM-based Speaker Similarity (SIM)~\cite{anastassiou2024seed}. To assess the distributional gap, we compute the Fréchet Audio Distance (FAD)~\cite{kilgour2018fr} and FD from PANNs~\cite{kong2020panns}.

\textbf{Human Evaluation.} Ten annotators rated 100 utterances per model on a 5-point scale for two dimensions: (1) NMOS (Naturalness), assessing acoustic fidelity, text-NV prosodic continuity and speaker consistency; (2) IMOS (Instruction Accuracy), evaluating precise NV execution without omissions, hallucinations and mispronunciations.

\subsubsection{Results and analysis}

\hspace*{\parindent}Table~\ref{tab:detailed_results} presents performance across Mandarin and English subsets. In terms of instruction alignment, NV-CV3 achieves the lowest PCER (27.69\%) and OCER (4.90\%) on the Mandarin single-label subset, demonstrating superior controllability with large-scale, diversified training data.

For acoustic fidelity, NV-CV3 maintains the strong timbre consistency of CV3, while Orpheus-TTS scores highest in DNSMOS perception. Table~\ref{tab:global_fad} shows NV-FlexiVoice achieves the lowest FAD and FD, indicating that its generated speech and NV events most closely match the real distribution.

Subjective evaluations support these results, with NV-CV3 achieving the highest NMOS and IMOS. Crucially, IMOS exhibits a significant negative Spearman correlation with PCER ($\rho = -0.65, p < 0.001$) and NMOS aligns with FD, thereby validating the reliability of NV-Bench's objective metrics.

\section{Conclusion}

We introduce \textbf{NV-Bench}, the first comprehensive benchmark for NV-capable TTS systems. Grounded in a functional taxonomy, the benchmark comprises 1,651 multi-lingual, in-the-wild utterances paired with human ground-truth, balanced across 14 categories into single- and multi-label subsets. NV-Bench evaluates acoustic fidelity and instruction alignment separately, thereby disentangling audio quality from paralinguistic controllability. Extensive experiments confirm that our objective metrics strongly correlate with human judgments, establishing NV-Bench as a standardized evaluation framework.

\section{Generative AI Use Disclosure}
We utilized LLM solely for the purpose of refining the clarity and grammar of the text. The authors reviewed and revised the output to ensure accuracy.

\bibliographystyle{IEEEtran}
\bibliography{mybib}

\end{document}